\documentclass[]{IEEEtran}
\usepackage{graphicx}
\usepackage{cite}
\usepackage[version=3]{mhchem}
\usepackage{amssymb}
\usepackage{color}
\usepackage{algorithm}
\usepackage{algpseudocode}
\usepackage{amsmath, multicol}

\newcommand{\R}{\mathbb{R}}

\newcommand{\Ugtrless}{%
	\mathrel{\kern0pt\mathop{\gtrless}\limits^{1}_{0}}%
}
\usepackage{textcomp}

\begin{document}
	\title{A Physical Channel Model for Wired Nano-Communication Networks}
	\author{Oussama~Abderrahmane~Dambri,~\IEEEmembership{Student Member,~IEEE,} and~Soumaya~Cherkaoui,~\IEEEmembership{Senior Member,~IEEE}
	}
	\maketitle
	\begin{abstract}
		
		In this paper, we propose a new end-to-end system for wired nano-communication networks using a self-assembled polymer. The self-assembly of a polymer creates a channel between the transmitter and the receiver in the form of a conductive nanowire that uses electrons as carriers of information. We derive the channel's analytical model and its master equation to study the dynamic process of the polymer self-assembly. We validate the analytical model with numerical and Monte-Carlo simulations. Then, we approximate the master equation by a one-dimensional Fokker-Planck equation and we solve this equation analytically and numerically. We formulate the expressions of the polymer elongation rate, its diffusion coefficient and the nullcline to study the distribution and the stability of the self-assembled nanowire. This study shows promising results for realizing stable polymer-based wired nanonetworks that can achieve high throughput.
		
	\end{abstract}
	
	\section{Introduction}
		\IEEEPARstart{N}{anomachines} are used in pharmaceutical and medical applications such as monitoring, drug delivery and real time chemical reactions detection. However, the capabilities of nanomachines are still very limited. This has prompted an interest in the design of nanonetworks that allow nanomachines to share information and to cooperate with each other \cite{akyildiz_internet_2010}. Nanonetworks call for a new networking paradigm that adapts traditional communication models to the nanoscale systems requirements. Two kinds of solutions are proposed in literature to create nanoteworks: first, using the electromagnetic waves in the Terahertz (THz) band, or second, using bio-inspired molecular communications.

	Using the THz band for electromagnetic nanonetworking is necessary given the nanoscale of the antennas\cite{akyildiz_electromagnetic_2010}. Communications in the THz frequency band suffer from scattering losses, molecular absorption and path loss \cite{doro_timing_2015, alsheikh_grid_2016, yao_tab-mac:_2016, jianling_high-throughput_2016, han_ma-adm:_2017, jadidi_tunable_2015, zhang_design_2016, hosseininejad_study_2017, zakrajsek_design_2017, zarrabi_wide_2017, wang_toward_2017, kim_statistical_2016, kokkoniemi_frequency_2015, han_three-dimensional_2017, he_stochastic_2017, piro_terahertz_2016, nafari_-chip_2017, guo_intra-body_2016}. Molecular communication is a promising bio-inspired solution to enable nanonetworks. Instead of using electromagnetic waves, molecules are used as wireless carriers of information between the transmitter and the receiver \cite{pierobon_physical_2010}. However, the achievable throughput with molecular communications is still very low and the delay is very high, despite the efforts reported in the literature to enhance both of them and to decrease the inter-symbol interference \cite{farsad_comprehensive_2016, arjmandi_ion_2016, mosayebi_type_2016, nakano_molecular_2017, chahibi_propagation_2016, deng_3d_2016, damrath_equivalent_2017, bicen_linear_2016, ahmadzadeh_stochastic_2017, tavakkoli_performance_2017, ardeshiri_performance_2017, tavakkoli_optimal_2017, tepekule_novel_2015, tepekule_isi_2015, kim_symbol_2014, akdeniz_optimal_2018, chang_adaptive_2018, assaf_influence_2017, dambri_MIMO_2019, noel_improving_2014, cho_effective_2017, dambri_performance_2019, dambri_enhancing_2018, einolghozati_networks_2016, unluturk_end--end_2017, barros_ca2+-signaling-based_2017, enomoto_design_2011}.
	
	\begin{figure}
		\centering
		\includegraphics[width=\linewidth]{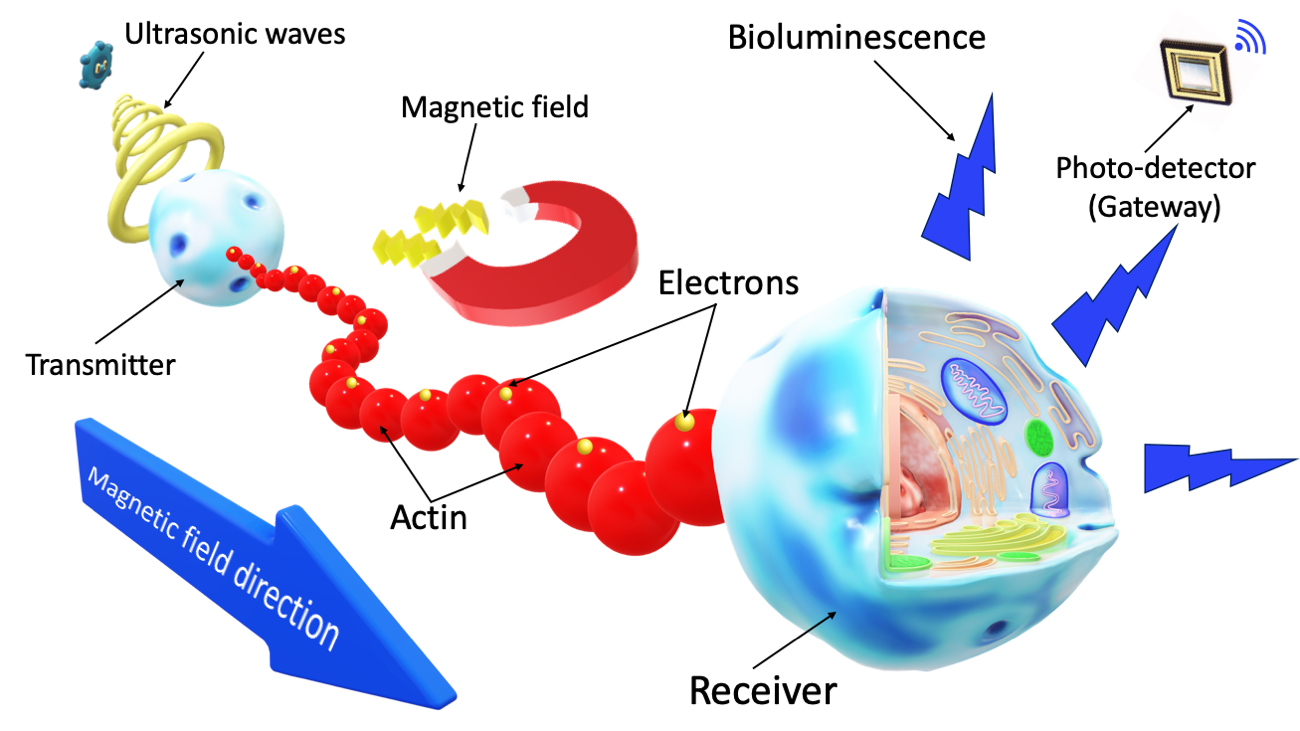}
		\caption{System design, where the transmitter contains ZnO matrix inside of it to transform ultrasonic waves into electricity. The self-assembled polymer guided by the magnetic field, connects the transmitter to the receiver and transports the electrons. When the receiver absorbs the electrons, it emits a bioluminescent light. The emitted light is detected by a photo-detector, which can play the role of a gateway.}
		\label{fig:1}
	\end{figure}
	
	In our previous work [1], we proposed the first attempt to use electrons as carriers of information in a wired system at the nanoscale. The proposed system is based on a polymer called "actin" \cite{lodish_actin_2000}, which self-assembles to construct a filament that plays the role of a conductive nanowire. Actin is a bi-globular protein with self-assembly capabilities, that is naturally used by biological systems such as the human body for cell skeletal maintenance, cell movement, cell division \cite{alberts_self-assembly_2002}. The direction of self-assembled actin filaments is random, however, experimental studies proved that we can guide actin filaments to a desired direction by using electric \cite{patolsky_actin-based_2004, Arsenault_2007, kaur_electric_2012} or magnetic fields \cite{kaur_low-intensity_2010}. The polymer-based system was shown to enable wired nano-communication networks  that are capable of very high throughput while presenting biocompatibility advantages [1]. In fact, the achievable throughput was demonstrated to reach mega bits per second \cite{dambri_toward_2019}, which is several orders of magnitude bigger than the throughput reported in the literature for other molecular communication systems \cite{farsad_comprehensive_2016, arjmandi_ion_2016, kim_symbol_2014, noel_improving_2014}. 
	
	The design of the end-to-end wired polymer-based system is shown in Fig. 1, where the transmitter contains a ZnO matrix to transform ultrasonic waves into electricity. The electrons are then sent through a self-assembled actin filament, which is guided by a magnetic field towards the receiver's direction. Finally, the receiver plays the role of a relay by absorbing the electrons and emitting bioluminescent light, which will be detected by a photo-sensor (gateway). The intensity of the light emitted by the receiver is dependent on the transmitted electrons' intensity. This property can be used as a modulation technique.

	This paper is an extension of the work proposed in [1], where we introduced the system concept and some preliminary simulation results. In this paper, we propose an analytical model to study the dynamic process of the polymer self-assembly, its elongation rate and its diffusion. We also validate the model with numerical and Monte-Carlo simulations. The main contributions of the paper are summarized as follows:

	\begin{enumerate}
		
		\item We derive a master equation of the communication channel to study the dynamic process of the polymer self-assembly, and we calculate the steady state of the self-assembly chemical reaction analytically and numerically.
		
		\item We validate the proposed analytical model with numerical and Monte-Carlo simulations and we present the simulation results in terms of the monomers' concentration changes over time. 
		
		\item We approximate the derived master equation by a 1D Fokker-Planck equation and we solve this equation  analytically by using a differential transform method, and numerically with Monte-Carlo simulations.
		
		\item We derive the nanowire elongation rate and its diffusion coefficient expressions, and we validate them with numerical simulations.
		
		\item We derive a nullcline expression to study the stability of the constructed self-assembled nanowire, and we provide an analysis using a phase plane graph.
		
	\end{enumerate}
	
	The rest of the paper is organized as follows. In section II, we summarize the state-of-the-art of the works proposed in the literature for electromagnetic and molecular communications, their advantages and disadvantages. In section III, we present an in-depth description of the proposed end-to-end wired polymer-based system design including a detailed view on the transmitter and the receiver. We then put the spotlight on the communication channel and study the dynamic assembly and distribution of the nanowire. In section IV, we derive the master equation of the nanowire self-assembly, and we calculate the steady state of the chemical reaction. We also derive the nullcline expression to study the stability of the self-assembled nanowire. Then, we approximate the derived master equation by a 1D Fokker-Planck equation. We solve the equation analytically by using a differential transform method and numerically with Monte-Carlo simulations. In section V, we discuss the obtained results of the analytical, numerical and Monte Carlo simulations. Finally, we present the conclusion in section VI.
	
	\section{State-Of-The-Art}
	
	The work on THz nanonetworks in literature can be categorized into three classes; a) Medium Access Control (MAC) protocol enhancement \cite{doro_timing_2015, alsheikh_grid_2016, yao_tab-mac:_2016, jianling_high-throughput_2016, han_ma-adm:_2017}, b) antenna enhancement designs\cite{jadidi_tunable_2015, zhang_design_2016, hosseininejad_study_2017, zakrajsek_design_2017, zarrabi_wide_2017, wang_toward_2017}  and c) channel modeling methods \cite{kim_statistical_2016, kokkoniemi_frequency_2015, han_three-dimensional_2017, he_stochastic_2017, piro_terahertz_2016, nafari_-chip_2017, guo_intra-body_2016}. In the first class, the researchers aim to enhance MAC protocols by designing energy-efficient nanonetworks \cite{doro_timing_2015, alsheikh_grid_2016} with high throughput and low delay \cite{yao_tab-mac:_2016, jianling_high-throughput_2016, han_ma-adm:_2017}. To enhance THz antennas, researchers use plasmonic waves \cite{jadidi_tunable_2015, zhang_design_2016, hosseininejad_study_2017} and take advantage of the black phosphorus and graphene's physical properties \cite{zakrajsek_design_2017, zarrabi_wide_2017, wang_toward_2017} to overcome transceivers' design problems. The last class studies the propagation of THz waves in different media, using 2D \cite{kim_statistical_2016, kokkoniemi_frequency_2015} and 3D models \cite{han_three-dimensional_2017, he_stochastic_2017}, by considering molecular absorption \cite{piro_terahertz_2016, nafari_-chip_2017} and scattering losses \cite{guo_intra-body_2016}. However, the molecular absorption of THz waves is very high, which drastically increases its path loss. The scattering and path losses are the main challenges that need to be tackled in order to capture and address the peculiarities of the THz band at nanoscale. Nanonetworks could be used to expand the capabilities of single nano-machines inside the human body for medical applications. However, using THz waves for such applications could be dangerous, because the vibration of water molecules increases when they absorb electromagnetic waves in such high frequency bands. The vibration raises the heat in the medium and can cause tissues' burn \cite{chopra_thz_2016}. Further studies are needed to investigate and ensure the safety of using THz nanonetworks inside the human body.

	Molecular communication is a promising bio-inspired solution to design nanonetwork systems. Instead of using electromagnetic waves, molecules are used as wireless carriers of information between the transmitter and the receiver\cite{farsad_comprehensive_2016}. From bacterial colonies to the human brain, molecular communication has been adopted by nature billions of years ago, and has demonstrated its efficiency at the nanoscale. This explains the interest of the research community in designing nanonetwork biosystems based on molecular communications for medical applications inside the human body. However, molecular communications have many challenges, the most important of which being its limited throughput, its high delay and the InterSymbol Interference (ISI) presence. After a previous transmission, some molecules remain in the medium because of their random walk, caused by thermal fluctuations at nanoscale. The molecules remaining in the medium interfere with the newly transmitted molecules, thus introducing errors at the receiver and affecting the reliability of communications.

	The most studied methods proposed for molecular communication are based on the principle of molecular diffusion, where molecules move randomly, because of the thermal fluctuations of the medium, until they reach the receiver. The work on molecular communication in the literature can be categorized into five classes; a) modulation techniques \cite{farsad_comprehensive_2016, arjmandi_ion_2016, mosayebi_type_2016, nakano_molecular_2017}, b) channel modeling studies \cite{chahibi_propagation_2016, deng_3d_2016, damrath_equivalent_2017, bicen_linear_2016, ahmadzadeh_stochastic_2017}, c) relay assistance methods \cite{tavakkoli_performance_2017, ardeshiri_performance_2017, tavakkoli_optimal_2017}, d) ISI avoidance \cite{tepekule_novel_2015, tepekule_isi_2015, kim_symbol_2014, akdeniz_optimal_2018, chang_adaptive_2018, assaf_influence_2017, dambri_MIMO_2019, noel_improving_2014, cho_effective_2017, dambri_performance_2019, dambri_enhancing_2018} and e) end-to-end communication system designs\cite{einolghozati_networks_2016, unluturk_end--end_2017, barros_ca2+-signaling-based_2017, enomoto_design_2011}. In the first class of works, the researchers propose adaptations of some of the techniques classically used with electromagnetic wave communications, to the new paradigm of molecular communications. These include aspects of modulation techniques, by controlling molecules' concentration (as signal amplitude), type (as signal frequency), and release time (as signal phase). \cite{farsad_comprehensive_2016}. Other researchers propose new ideas such as; using ion protein channels to control molecules release \cite{arjmandi_ion_2016}, a ratio shift between two types of molecules \cite{mosayebi_type_2016} or the dynamic properties of propagation patterns in molecules' concentration \cite{nakano_molecular_2017}. In the second class of works, the researchers modeled molecular communication channels to study the dynamic distribution of information in the medium \cite{chahibi_propagation_2016, deng_3d_2016}. They use discrete-time channels \cite{damrath_equivalent_2017}, or continuous stochastic models for constant \cite{bicen_linear_2016} and mobile transmitters and receivers \cite{ahmadzadeh_stochastic_2017}. To minimize error probabilities and optimize molecular communication performance, the works in the third class propose using a relay-assisted diffusion between transmitters and receivers \cite{tavakkoli_performance_2017, ardeshiri_performance_2017, tavakkoli_optimal_2017}. Avoiding ISI is a main challenge of molecular communication. In the fourth class of works, two types of solutions are proposed to avoid ISI: passive and active solutions. Passive solutions use pre-equalization methods \cite{tepekule_novel_2015, tepekule_isi_2015}, or optimize symbol times and detection thresholds \cite{kim_symbol_2014, akdeniz_optimal_2018, chang_adaptive_2018}. Active solutions propose removing the molecules physically from the medium to avoid ISI, by using neighboring receivers  \cite{assaf_influence_2017, dambri_MIMO_2019}, enzymes \cite{noel_improving_2014, cho_effective_2017} or photolysis reactions \cite{dambri_performance_2019, dambri_enhancing_2018}. In the last class of works, inspired by nature, the researchers propose more complex biosystem nanonetworks such as bacteria colonies \cite{einolghozati_networks_2016}, plant pheromones \cite{unluturk_end--end_2017}, Ca$^{2+}$ signaling in the cells \cite{barros_ca2+-signaling-based_2017, bicen_linear_2016} and using molecular motors\cite{enomoto_design_2011, chahibi_propagation_2016}. The last system uses microtubule polymers as a road for kinesin motors to walk on, and transport a cargo of information from the transmitter to the receiver. Despite the efforts reported in the literature to enhance molecular communication performance, the achievable throughput still very low (tens of bits per second), with very high delay (minutes, hours). In the present work, we use a polymer of the kinesin motor reported in \cite{enomoto_design_2011}  as an electric nanowire instead of a road, and we show that we can design nanonetworks using this kind of nanowire which can
	achieve high throughput (Mbits per second).

	\section{System Design}
	
	The design of a flexible wired nano-communication network is a new promising solution to be used in pharmaceutical and medical applications that can provide a very high throughput. The proposed polymer-based nano-communication system has the potential to be implemented noninvasively inside the human body due to its tiny size, and the fact that electrons generation can be controllable from the outside with ultrasonic waves, as explained in Fig. 1. The proposed end-to-end system contains a transmitter with a ZnO matrix, a receiver with photo-proteins and a photo-sensor as the gateway. The monomers diffuse randomly in the medium and their self-assembly constructs a nanowire which links the transmitter to the receiver.
	
	\subsection{Transmitter}
	
	TThe innovative approach proposed in \cite{wang_top_2010} that converts mechanical energy into electricity at nanoscale by using piezoelectric Zinc Oxide (ZnO) nanowire matrix is used in our transmitter to generate electrons. The piezoelectric potential is created by the polarization of the ions in some solid materials such as crystals and ceramics, or biological matter as  DNA and some proteins, when subjected to strain. The authors of \cite{wang_top_2010}  used the unique coupling of semiconducting and piezoelectric dual ZnO properties and a Schottky barrier between the metal tip and the nanowire to create a DC nano-generator. The transmitter in the proposed system uses this DC nano-generator to convert mechanical vibration into electricity. The mechanical vibration is generated with ultrasonic waves. Pulses of ultrasonic waves are converted into pulses of electrons at the transmitter, which sends them through the assembled nanowire. To modulate the information at the transmitter, we can use the amplitude and the frequency of the ultrasonic waves.
	
	\begin{figure}
		\centering
		\includegraphics[width=\linewidth]{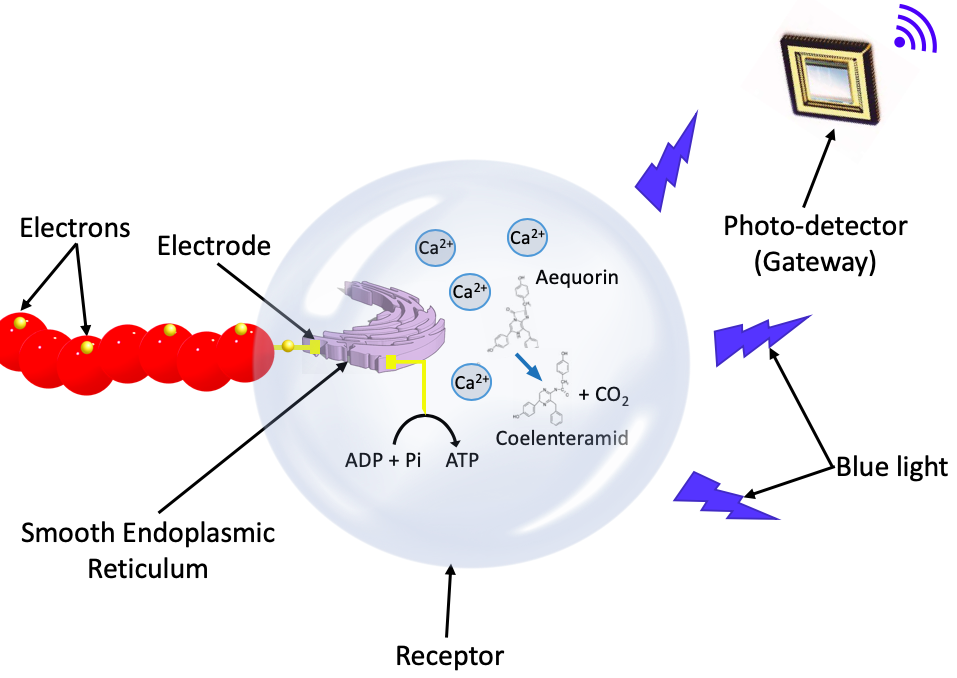}
		\caption{Receiver design that uses electrons to generate bioluminescent blue light.}
		\label{fig:2}
	\end{figure}
	
	\begin{figure}
		\centering
		\includegraphics[width=\linewidth]{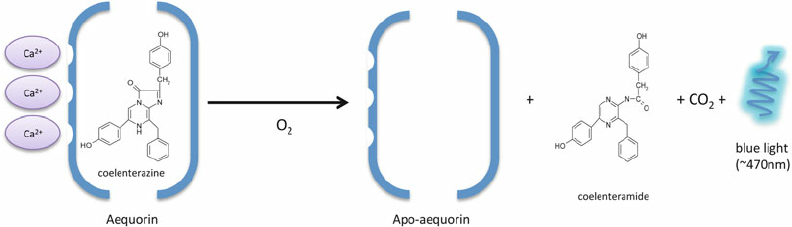}
		\caption{Bioluminescent reaction that uses Aequorin, which in the presence of Ca$^{2+}$ ions, it generates blue light \cite{badr_bioluminescence_2014}. }
		\label{fig:3}
	\end{figure}
	
	\subsection{Receiver}
	
	The electrons sent through the assembled nanowire will be absorbed by the receiver. However, it is extremely difficult for a user to extract the sent information from the received electrons at nanoscale because of the quantum trade-off between information and uncertainty. Therefore, our proposed receiver is designed to use the received electrons to provoke chemical reactions that generate bioluminescent light, which makes the extraction of information easier. Bioluminescence is a chemical emission of light by living organisms using light-emitting molecules (photo-proteins) and enzymes. Bioluminescence is a promising solution to make the shift from nanoscale to the micro and macro scales in our proposed receiver, as shown in Fig. 2. There are several photo-proteins in biological systems, the most studied and most famous one is \textit{Luciferin} with its enzyme \textit{Luciferase}. \textit{Luciferin} can be found in fireflies and deep-sea fishes. The oxidation of \textit{Luciferin} is catalyzed by \textit{Luciferase}, and the resulting excited intermediate state emits light upon decaying to its ground state \cite{Aaron_2019}. In our proposed system, we use another photo-protein called \textit{Aequorin} which is constructed by the jellyfish \textit{Aequorea Victoria} that can be found in North America and the Pacific Ocean \cite{shimomura_short_1995}. In the presence of Ca$^{2+}$, the \textit{Aequorin} oxidation reaction described in Fig. 3 is activated, which generates blue light with 470 nm wavelength \cite{badr_bioluminescence_2014}.  
	
	The proposed receiver contains high concentrations of \textit{Aequorin}, and a Smooth Endoplasmic Reticulum (SER), which functions as a Ca$^{2+}$ ions  storage in living cells \cite{koch_endoplasmic_1990}. To avoid the absorption of electrons by the receiver's surface, we propose to build the receiver with a transparent insulating membrane. When the assembled nanowire reaches the receiver, it binds to one of the  monomers already anchored to the receiver's surface with electrodes, which creates a passage of electrons through the insulating membrane. The absorbed electrons excite the SER, as shown in Fig. 2, which causes the secretion of Ca$^{2+}$ ions. The \textit{Aequorin} inside the receiver activates in the presence of Ca$^{2+}$ ions and emits blue light, which is detected by a photo-sensor as a gateway. When the emission of electrons ceases, the secretion of Ca$^{2+}$ ions stops, and SER absorbs all Ca$^{2+}$ ions inside the receiver as a sponge. Without Ca$^{2+}$ ions, \textit{Aequorin} becomes inactivated and stops emitting blue light. The intensity of the blue light emission from the receiver depends on the emitted electrons intensity (at the transmitter), which itself depends on the emitted ultrasonic waves intensity (from outside the body). 
	
	Several challenges need to be tackled in our proposed receiver design. First, we need to calculate the number of electrons needed to excite the SER. Then, we determine the relation between the number of the absorbed electrons and the concentration of Ca$^{2+}$ ions secreted by the SER. Finally, we need to establish a relation between the bioluminescent light intensity detected by the photo-sensor and the intensity of the current sent from the transmitter. An experimental study is envisioned to be proposed in our future work.
	
	\subsection{Channel}
	
	The cytoskeleton is an essential complex of interlinking filaments for the living cells shape, movements and division  \cite{alberts_self-assembly_2002}. Three types of filaments construct the cytoskeleton complex, depends on the assembled protein that constructs them. Microfilaments are constructed with assembled actin, microtubules are constructed with assembled tubulin and the intermediate filaments are constructed with keratin, vimentin, lamin or desmin. The choice of actin as our polymer in this study to construct the proposed nanowire is due to the fact that microfilaments constructed by actin are more flexible, an easily controllable, compared to microtubules and intermediate filaments. Actin is one of the most studied and most abundant proteins, comprising 10\% of muscle cells total proteins and around 5\% in the other cells  \cite{lodish_actin_2000}, which makes it an ideal candidate to construct our proposed nanowire. Actin filaments have a high electrical conductivity, as proven in the experimental studies in  \cite{patolsky_actin-based_2004} and  \cite{hunley_multi-scale_2018}. The authors in  \cite{patolsky_actin-based_2004} proposed a metallic actin-based nanowire, where they labelled globular actin (G-actin) with gold nanoparticles to increase the electrical conductivity of actin filaments (F-actin). The study showed that the electrical conductivity of a metallic actin-based nanowire can reach 25 $\mu$A for 0.8 mV potential, which is very high at nanoscale. Fig. 4 shows a High Resolution Scanning Electron Microscopy (HRSEM) image of the studied metallic nanowire between two electrodes  \cite{patolsky_actin-based_2004}. The authors in  \cite{hunley_multi-scale_2018} and  \cite{tuszynski_ionic_2004} studied the electrical impulses and ionic waves propagating along actin filaments in both intracellular and \textit{in vitro} conditions. The results of the studies revealed the existence of electrical signal impulses and ionic waves propagating through intracellular actin filaments in the form of solitons at a speed reaching 0.03 m/s. However, the polymerization process of G-actin monomers to form F-actin filament is random, because of the thermal fluctuations in the medium. One of the main challenges to construct a nanowire in a randomly diffusive medium is to guide its assembly to a desired direction.
	
	\begin{figure}
		\centering
		\includegraphics[width=\linewidth]{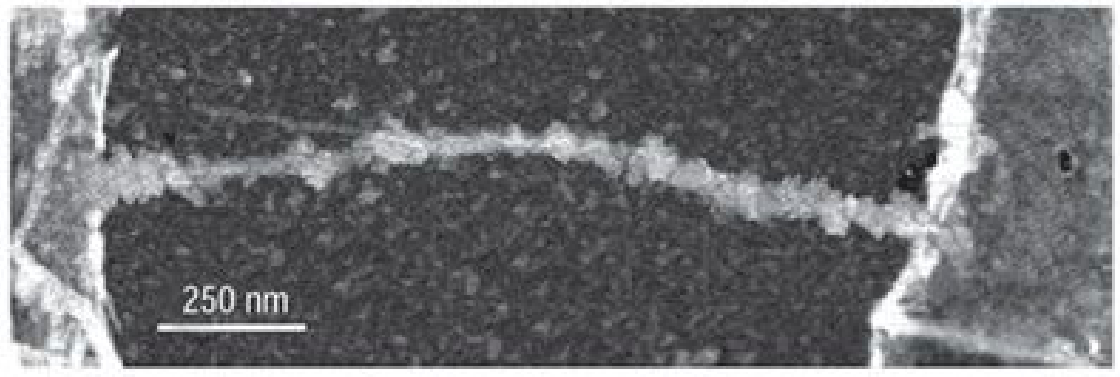}
		\caption{High Resolution Scanning Electron Microscopy (HRSEM) image of a metallic actin-based nanowire between two electrodes  \cite{patolsky_actin-based_2004}.}
		\label{fig:4}
	\end{figure}
	
	To guide the direction of actin filaments assembly, two solutions are proposed in the literature, either by using electric \cite{patolsky_actin-based_2004, Arsenault_2007, kaur_electric_2012} or magnetic field \cite{kaur_low-intensity_2010}. The experimental study in \cite{Arsenault_2007} has shown that by applying an electric field to the actin filaments, they align themselves parallel to the field lines. The study proved that as the electric field intensity increases, the variance in actin displacement decreases, which affect the filament's behavior and cause its alignment. The authors in \cite{kaur_electric_2012} also used AC electric field to guide the direction of actin assembly. The same authors proposed in \cite{kaur_low-intensity_2010} a new idea to align assembled actin filaments by using a magnet bar with a low-intensity magnetic field (22 mT). The study has shown that the majority of F-actin got permanently aligned towards the magnetic lines. The authors concluded that a magnetic field can be safely used to permanently orient and guide the alignment of F-actin towards a desired direction. Because using AC electric field inside the human body is dangerous, our proposed system shown in Fig. 1 is designed to guide the actin assembly direction by using a magnetic field. The path of the transmitted electrons through the proposed nanowire is circular because of the helicoidal shape of actin filaments.
	
	In the next section, we will derive an analytical model to study the dynamic process of the actin nanowire self-assembly. We also calculate the steady state of the chemical reaction, and we derive a nullcline expression to study the nanowire stability. 
	
	\section{Channel Model}
	
	Actin proteins exist under two forms, monomeric (G-actin), and filamentous (F-actin). The self-assembly of actin monomers creates actin filaments, and this polymerization reaction involves three steps: nucleation, elongation and steady state \cite{pollard_direct_1981}. The first step consists in the construction of an actin nucleation core, which is a three-actin monomers protein. This nucleation core is energetically unfavorable, unless it binds with another actin monomer, which leads to an energetically favorable elongation reaction \cite{liu_insertions_2013}. The second step is where the polymerization and depolymerization change the rate of actin filaments elongation. The polymerization and depolymerization  take place on both sides of actin filaments as shown in Fig. 5, but with different rates. The faster side is called the pointed end, and the slower side is called the barbed end. The final step occurs when the addition and dissociation of actin monomers to the filament sides are balanced and a steady state is reached. 
	
	\subsection{Chemical Reaction}
	
	In this paper, the studied polymerization of the actin nanowire does not consider the nucleation step because we assume that nucleation cores are already anchored at the transmitter's surface, as shown in Fig. 5. When an actin monomer binds with one of the anchored nucleation cores, the elongation of the nanowire is triggered, thus, the studied reaction in this paper starts directly with the elongation step. To simplify the complex polymerization reaction of the actin nanowire, we consider fixed reaction rates. We also consider a nanowire $N_i$ consisting of $\textit{i}$ actin monomers. When a monomer $N_1$ binds with the complex $N_i$, it becomes $N_{i+1}$, and a monomer dissociation gives $N_{i-1}$. Knowing that a nucleation core contains 3 monomers, the nanowire polymerization in this study starts with $N_4$. The pointed end is written as \cite{pollard_direct_1981}: 
	\begin{equation}
	N_i + N_1  \ce{<=>[\ce{$k_{1}^p$}][\ce{$k_{-1}^p$}]} N_{i+1},
	\label{eq:1}
	\end{equation} and for the barbed end:
	\begin{equation}
	N_i + N_1  \ce{<=>[\ce{$k_{1}^b$}][\ce{$k_{-1}^b$}]} N_{i+1},
	\label{eq:2}
	\end{equation} where $k_{1}^p$ and $k_{-1}^p$ are the polymerization and depolymerization rates for the pointed end, $k_{1}^b$ and $k_{-1}^b$ are the rates of the barbed end. $N_1$ is the actin monomer and $N_i$ is the actin nanowire, where $\textit{i} \in \mathbb{N}^*$ and 4 $\leq \textit{i} \leq l$. $l$ is the maximum number of actin monomers in the constructed actin nanowire.
	
	However, the anchored nucleation core at the transmitter's surface fixes the barbed end of the actin nanowire as shown in Fig. 5, and there will be no addition nor dissociation from that side. Therefore, this study uses only the reaction in (1), and to simplify the notations, we write $k_{+}$ and $k_{-}$ instead of $k_{1}^p$ and $k_{-1}^p$ respectively. The equation (1) is described by a differential equation as follows:
	\begin{equation}
	\frac{d}{dt} n(t)= - 2 k_{+} N_i^2 + k_{-},
	\label{eq:3}
	\end{equation} and the elongation of the actin nanowire is described as:
	\begin{equation}
	\frac{d}{dt}a(t)= -k_{-} + 2 k_{+} N_i^2,
	\label{eq:4}
	\end{equation} where $\frac{d}{dt}n(t)$ is the change of the monomers concentration in the medium with time. $\frac{d}{dt}a(t)$ is the change of the actin nanowire elongation with time.
	
	\begin{figure}
		\centering
		\includegraphics[width=\linewidth]{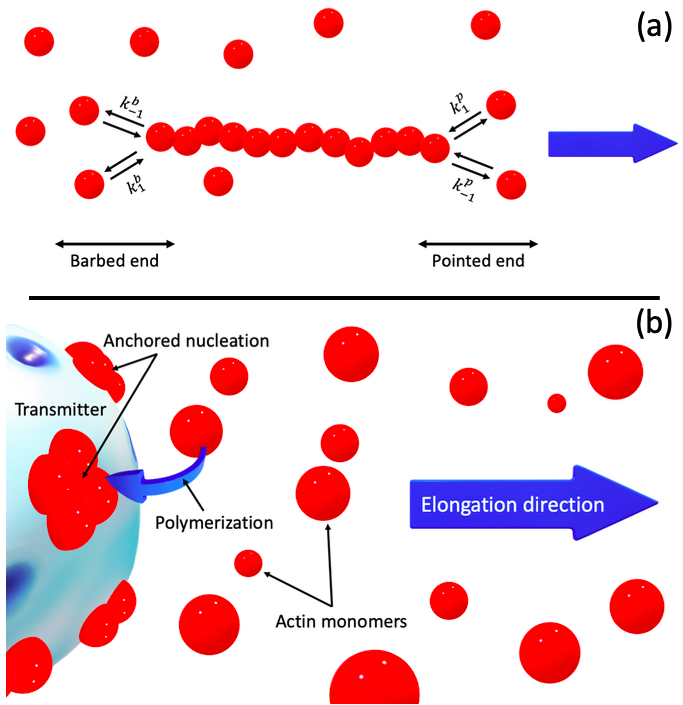}
		\caption{Polymerization and depolymerization of actin nanowire. (a) The fast pointed end and the slow barbed end of the actin self-assembly. (b) The nucleation already anchored in all the surface of the transmitter to quickly activate the actin elongation.}
		\label{fig:5}
	\end{figure}
	
	\subsection{Markov Process Model}
	
	The proposed nanowire channel is nucleation-limited because every G-actin monomer is unpopulated at most times, which makes their binding random. Thus, the actin polymerization must be treated as a stochastic system \cite{betz_stochastic_2009}. If the conditional probability distribution of a stochastic process's future states depends only upon the present state and not on the events that preceded it, the process is called a Markov process. Our proposed channel is formulated as a Markov chain, which is a type of Markov process that has either discrete state space or discrete time index set \cite{wilkinson_stochastic_2011}. Before deriving the chemical master equation of the proposed channel from the Markov process model, we need first to determine the reaction propensities. A reaction propensity tells us how likely a reaction is to occur per unit time \cite{ingalls_mathematical_2013}. If we assume that the reaction events are independent and that at most one reaction event between two actin monomers can occur at a time, then, the propensity of this bimolecular polymerization reaction is\cite{ingalls_mathematical_2013}:
	\begin{equation}
	n = \frac{k_+N(N-1)}{2},
	\label{eq:5}
	\end{equation} where $N$ is the number of actin monomers in the medium. The propensity of the depolymerization reaction is $k_-$. The reaction stoichiometries $s_i$ are $[1]$ or $[-1]$, which means that only one monomer is added or dissociated at a time. The polymerization that leads to the nanowire elongation is a Markov chain that changes the system, at each reaction, from a state $N$ to a state $N+1$. In our stochastic model, we assume a small-time increment, $dt$and we assume that, at each time interval [$t$, $t+dt$], the probability that a reaction occurs is the product of the interval length and the reaction propensity $n_i (N(t))dt$ \cite{ingalls_mathematical_2013}. Thus, the probability that no reaction occurs in the time interval is 1$- \sum_{i}^{l} n_i (N(t))dt$, where $i$ represents all the occurring reactions in the channel. 
	
	Let's consider $P(N, t)$ the probability that the channel is in the state $N$ at a time $t$, and is dependent on the initial conditions. If we know the distribution of $P(N, t)$ at time $t$, we can describe the distribution at an incremented time $t+dt$ as:
	\begin{equation}
	\begin{split}
	P(N, t+dt) = P(N, t) \bigg (1- \sum_{i}^{l} n_i (N)dt\bigg ) + \sum_{i}^{l} \\ P (N-s_i, t) n_i (N-s_i) dt,
	\end{split} 
	\label{eq:6}
	\end{equation}
	 where $s_i$ represents the reaction stoichiometries. The first half of the equation represents the probability of no reactions occurring. The second half represents the probability of the reactions firing, while in a state ($N-s_i$). By substituting the reaction propensities in the equation (6) we write: 
	 
	\begin{equation}
	\begin{split}
	 \text{\footnotesize $ P(N, t+dt) = P(N, t)(1-  (k_+ N(N-1)/2)dt) + P(N-1, t)$} \\ \text{\footnotesize $  (k_+(N-1)(N-2)/2)dt + P(N+1, t)k_-dt $},
	\label{eq:7}
	\end{split} 
	\end{equation}
	
	This equation is called the probability balance, and we can use it to derive our chemical master equation by substituting $P(N,t)$ from each side of equation (7), which gives:
	\begin{equation}
	\begin{split}
	\text{\footnotesize $ P(N, t+dt)-P(N, t) = -P(N, t)(k_+ N(N-1)/2)dt + P (N-1, t)  $}  \\ \text{\footnotesize $ (k_+(N-1)(N-2)/2)dt + P(N+1, t)k_-dt $},
	\label{eq:8}
	\end{split} 
	\end{equation} 
	
	This chemical master equation can be written as a differential equation, and we write: 
	\begin{equation}
	\begin{split}
	\text{\footnotesize $ \frac {d}{dt} P(N, t) = -P(N, t)(k_+ N(N-1)/2)dt + P(N-1, t) $}  \\ \text{\footnotesize $(k_+(N-1)(N-2)/2)dt + P(N+1, t)k_-dt $},
	\label{eq:9}
	\end{split} 
	\end{equation} 
	
	Our derived chemical master equation can also be written as an iterative probability differential equation by putting $P(N,t)$ as $P_i (t)$ and we write:
	\begin{equation}
	\frac {d}{dt} P(N, t) = -(k_+M+k_-)P_i(t) + k_+M P_{i-1}(t)+ k_-P_{i+1},
	\label{eq:10}
	\end{equation} where $M$=$N$($N-1$)/2. The master equation in (10) is a finite system of differential equations that describe the time-varying of the proposed actin nanowire probability distribution, and it can be written in a general form as:
	
	\begin{equation}
	\frac {d}{dt} P_i(t) = \sum_{i}^{l} \bigg (W_{i, i+1}P_{i+1}(t)+W_{i+1, i}P_i(t)\bigg ),
	\label{eq:11}
	\end{equation} where $W_(i, i+1)$ and $W_( i+1, i)$ are the transition probabilities. The exact solution of the master equations for bimolecular reactions are complex and hardly obtained. Several works have been proposed to find the solution of some bimolecular reactions' master equation \cite{laurenzi_analytical_2000, lee_analytical_2012}. Other works use the Gillespie algorithms to run stochastic Monte-Carlo simulations \cite{gillespie_rigorous_1992, ingalls_mathematical_2013, wilkinson_stochastic_2011}. The disadvantage of this method is that it needs a lot of trajectories in order to estimate accurately the master equation's solution \cite{szekely_stochastic_2014}. The approximation of the master equation by a 1D Fokker-Planck equation is much more efficient computationally and gives an accurate solution. In the last subsection, we will approximate the derived master equation by a 1D Fokker-Planck equation, and we solve it analytically and numerically.
	
	\subsection{Reaction Steady State}
	
	The actin self-assembly reaches a steady state when the addition and dissociation of actin monomers is balanced, which means that the change rate of actin concentration is zero and we can write: 
	\begin{equation}
	- 2k_{+} N_{ss}^2 + k_{-} = 0,
	\label{eq:12}
	\end{equation} where $N_{ss}$ is the actin monomers steady state concentration, which is equal to an equilibrium constant $K=\sqrt {k_-/2k_+}$ (critical concentration) for polymerization \cite{pollard_direct_1981}. We expect the solution of the differential equation in (3) to approach the steady state value exponentially, and we can derive an explicit description of the time-varying concentration as:
	\begin{equation}
	n(t)= (N_0-K)e^{-2k_+t}+ K,
	\label{eq:13}
	\end{equation} where $N_0$ is the initial concentration of actin monomers, and $K$ is the equilibrium constant. The solution indicates that when $t \rightarrow \infty$ the concentration of actin monomers decays  exponentially until it reaches the critical concentration of polymerization $K$. Phase plane analysis is another approach to study the steady state of a dynamic system. Instead of plotting the concentration of actin monomers as functions of time, the phase plane plots the decaying concentration of actin monomers in the medium $n(t)$ against the concentration of actin monomers constructing the nanowire $a(t)$. Phase plane analysis shows the trajectories that the concentrations take starting from initial conditions and converging to the steady state. The total points ($n_i$, $a_i$) where the trajectories of the phase plane change their direction constitutes the channel's nullclines. The determination of these nullclines analytically is not always possible, because the function $g$($n_i$, $a_i$) is usually nonlinear and so may not be solvable except via numerical simulations. However, it is possible to determine the analytical expression of the nullcline in our proposed linear system, which represents the function $g$($n_i$, $a_i$)  and we write (3) as:
	\begin{equation}
	0= - 2 k_{+} n_i^2 + k_{-}a_i,
	\label{eq:14}
	\end{equation} 
	
	Then, the nullcline expression can simply be written as:
	\begin{equation}
	n_i= K\sqrt {a_i},
	\label{eq:15}
	\end{equation} 
	
	We notice that the nullcline expression of our proposed channel depends on the ratio $k_-/2k_+$. When $k_+ > k_-$, the actin nanowire elongates with polymerization, and when $k_+ < k_-$, the actin nanowire dissociates with depolymerization.
	
	\subsection{Channel's Stability}
	
	The dynamic process of the biological systems do not often diverge or follow a chaotic behavior. The biochemical systems either converge to a steady state, or converge to a sustained periodic oscillations \cite{ingalls_mathematical_2013}. The proposed actin nanowire self-assembly also converges to a steady state, and to see if that steady state is stable of unstable, we need to calculate the eigenvalues of the channel's Jacobian matrix. The stability of the channel depends on the eignevalues signs. If all eigenvalues are negative, then the steady state of the system is stable. If one of the system's eigenvalues is positive, then the system is unstable. The differential equations described in (3) and (4) can be written as a square Jacobian matrix as follows:
	\begin{equation}
	J= \begin{bmatrix}
	-4k_+N(t)     & 0 \\
	\ \ 4k_+N(t)       & 0 
	\end{bmatrix}
	\label{eq:16}
	\end{equation} 
	
	The eignevalues of this Jacobian matrix represent the roots of the quadratic equation:
	\begin{equation}
	\lambda^2+4k_+N(t) \lambda = 0,
	\label{eq:17}
	\end{equation} 
	which gives a zero eigenvalue $\lambda_1=0$ and a negative eigenvalue $\lambda_2 = - 4k_+N(t)$. The case where one of the eigenvalues is zero happens only if the system have more than one equilibrium point as confirmed in the nullcline expression (14). In this case, we can say that the system is stable, but not asymptotically stable, where the stability is determined by the nonlinear terms of the system's equations \cite{roussel_2019}. Therefore, we also studied the stability of the proposed nanowire empirically by observing the results of our 3D stochastic simulations, and comparing them with experimental studies in the literature. The non-linearity of the proposed channel is influenced by the enzyme concentrations, the assembled nanowire length and the magnetic field that guides the direction of the nanowire self-assembly, and we write:
	\begin{equation}
	Stability = \frac{M \times E}{L}
	\label{eq:18}
	\end{equation} 
	Where $M$ is the intensity of the magnetic field $E$ is the enzyme concentration, and $L$ is the length of the nanowire. The Fig. 6 shows the influence of the enzyme concentration and the length on the nanowire stability with three different magnetic field intensity values. The more the intensity of the magnetic field, the less fluctuations in the medium, and the more stable is the proposed nanowire.
	
	\subsection{Fokker-Planck Equation}
	
	The one-dimensional Fokker-Planck equation emerges in the biological, chemical and physical sciences as an excellent approximation to the master equations, because of its elegant mathematical properties \cite{kampen_stochastic_2007}. The 1D Fokker-Planck in its general form can be written as \cite{kampen_stochastic_2007}:
	
	\begin{equation}
	\frac{\partial u(x, t)}{\partial t} = \bigg [-\frac{\partial }{\partial x} A(x, t) + \frac{\partial^2 }{\partial x^2} B(x, t) \bigg ]u,
	\label{eq:19}
	\end{equation} where $u(x,   t)$ is an unknown function that can represent the probability density in our study. $B(x,t)>0$ is the diffusion coefficient, and  $A(x,t)$ is the drift coefficient, with the following initial condition $u(x,   0)=f(x)$, $x \in \R$. To derive the Fokker-Planck equation, we need to discretize the chemical master equation into small jumps. The transition probabilities $W$ in (11) change from a state $N$ to a state $N+1$ with a small size jump, and we write \cite{kampen_stochastic_2007}:
	
	\begin{equation}
	W(N|N+1)=W(N+1; \delta), \: \; \; \;  \;  \delta=(N+1)-N,
	\label{eq:20}
	\end{equation} where $\delta$ represents the distance between two neighboring actin monomers. By replacing the probability $P_i (t)$ with a probability density $p(x_i,t)$, the master equation, then, is written as \cite{kampen_stochastic_2007}:
	
	\begin{equation}
	\text{\footnotesize $  \frac {\partial p(x_i,t)}{\partial t} = \int W(x_i-\delta; \delta)p(x_i-\delta, t)d\delta - p(x_i, t) \int W(x_i; -\delta)d\delta, $}
	\label{eq:21}
	\end{equation} where $x_i=i\times \delta$ is the position of the $i^{th}$ monomer in the actin filament, by assuming that only small jumps occur, we calculate the integral by means of a Tylor expansion up to second order and we write: 
	
	\begin{figure}
		\centering
		\includegraphics[width=\linewidth]{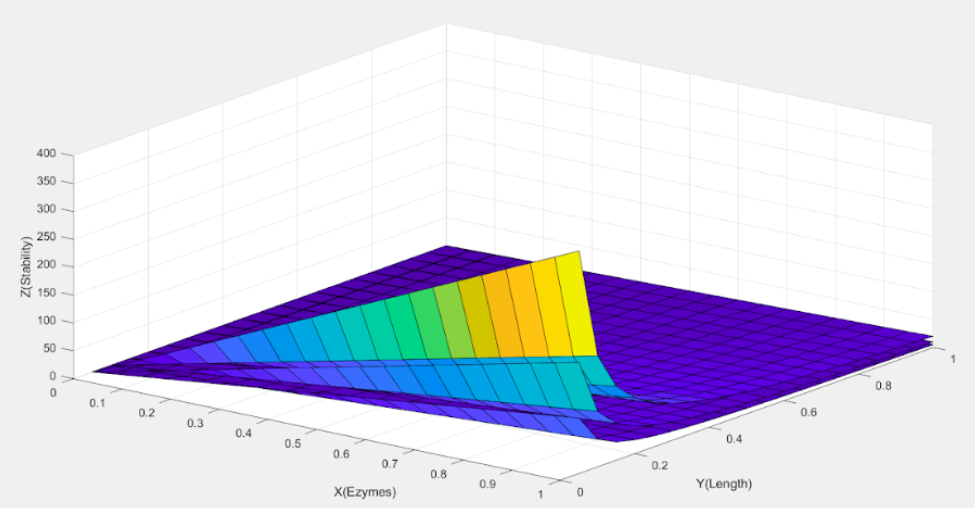}
		\caption{Enzyme and length effects on the nanowire stability using three intensity values of the magnetic field [1].}
		\label{fig:6}
	\end{figure}
	
	\begin{equation}
	\frac {\partial p(x,t)}{\partial t} \approx -(k_+N-k_-)\delta \frac {\partial p(x,t)}{\partial x} + \frac {k_+N+k_-}{2} \delta^2 \frac {\partial^2 p(x,t)}{\partial x^2}.
	\label{eq:22}
	\end{equation} 
	
	The eq. (22) is a special case of the Fokker-Planck equation known as the backward Kolmogorov equation. By comparison with eq. (19), we extract the elongation rate coefficient of the actin nanowire $E=(k_+ N_0-k_- )\delta$ , and its distribution coefficient $D=\frac {k_+N+k_-}{2} \delta^2$. We note that the variance of the nanowire length is time-dependent $\sigma^2=2Dt$. We use two methods to solve the eq. (22): numerically with simulations, and analytically by using Differential Transform Method (DMT) as in \cite{hesam_analytical_2012}. DMT constructs analytical solutions in the form of polynomials based on the Tylor series expansion. The advantage of this method is that it calculates solutions by means of an iterative procedure, which reduces the computational burden. The differential transform of a function $f(x, y)$, is \cite{hesam_analytical_2012}:
	
	\begin{equation}
	F(k, h)= \frac {1}{k!h!} \bigg [\frac {\partial^{(k+h)}f(x, y)}{\partial x^k \partial y^h }  \bigg ]_{(x=x_0, y=y_0)}, 
	\label{eq:23}
	\end{equation} 
	
	The inverse differential transform function of $F(k, h)$ for finite series is expressed as follows \cite{hesam_analytical_2012}:
	
	\begin{equation}
	f(x, y)= \sum_k^n \sum_h^m W(k, h)x^k y^h,
	\label{eq:24}
	\end{equation} 
	
	We use the fundamental mathematical operations performed by the equations (23) and (24), which are presented in Table. 1 in \cite{hesam_analytical_2012}. We determined an exact analytical solution of (22), which satisfy the boundary conditions of the channel, where $x_0$ is the surface of the transmitter, and $x=l$ is the surface of the receiver and we write the probability distribution of the actin nanowire as:
	
	\begin{equation}
	p(x, t)= \frac {1}{\sqrt{\pi Dt}} \exp \bigg (\frac {(x-x_0-E)^2}{2Dt}\bigg ).
	\label{eq:25}
	\end{equation} where $x$ is the position of monomers along the actin nanowire, $E$ is the elongation rate coefficient and $D$ is the distribution coefficient. We assumed that the initial concentration of the actin monomers $N_0$ is constant and that their initial distribution $p(x_0,0)$  is a Gaussian distribution, because of the random movements of actin monomers in the medium. We validated the analytical solution by using PDEPE function in MATLAB to calculate the Fokker-Planck equation numerically.

	\section{Numerical Results}
	
	Depending on the phosphorylation state of actin monomers, two types are distinguished; ATP-actin and ADP-actin, and each one of them has different reaction rates. Polymerization and depolymerization rates can also be influenced by other parameters namely; the medium viscosity and enzyme concentration, which explains the diversity of the rates values in the literature. In this study, we ignored the phosphorylation state of actin monomers, and we calculated the average of the reaction rates in the literature for the two monomer types using the viscosity of the human blood. The polymerization and depolymerization rates used in this study are approximated to $k_+\approx$ 0.979 $\mu M^{-1}s^{-1}$ and $k_- \approx$ 0.166 $s^{-1}$ respectively. The initial concentration of actin monomers in our simulations is N0 = 1000, and their initial distribution is a Gaussian because of their random movement in the medium. By assuming that the diameter of the transmitter and the receiver is 1 $\mu m$ and the distance between the surface of the transmitter and the surface of the receiver is 10 $\mu m$, we take the boundary conditions as $x_0 =$ 1 $\mu m$ and $x_l =$ 10 $\mu m$. The diameter of each actin monomer is approximately 5.5 nm \cite{tuszynski_ionic_2004}, thus, we take the distance between two neighboring actin monomers as the sum of their diameters namely; $\delta =$ 11 $nm$. 
	
	\begin{figure}
		\centering
		\includegraphics[width=\linewidth]{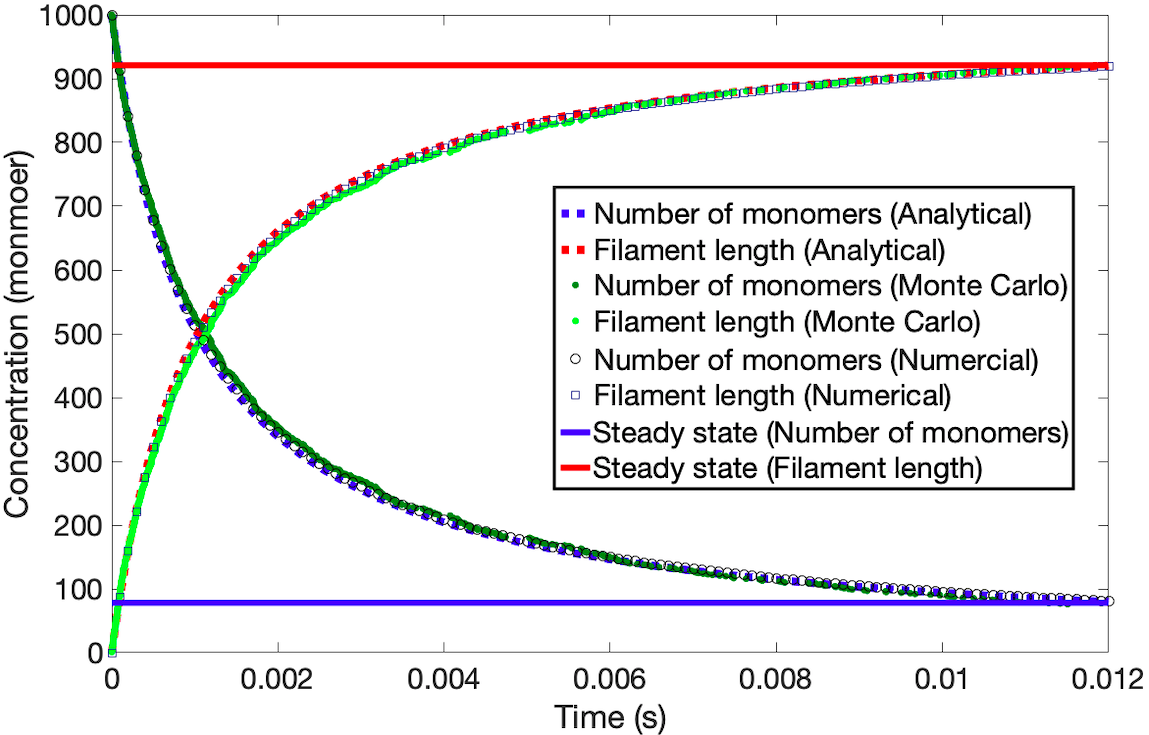}
		\caption{Time-varying behavior and steady state of actin nanowire formation.}
		\label{fig:7}
	\end{figure}
	
	\begin{figure}
		\centering
		\includegraphics[width=\linewidth]{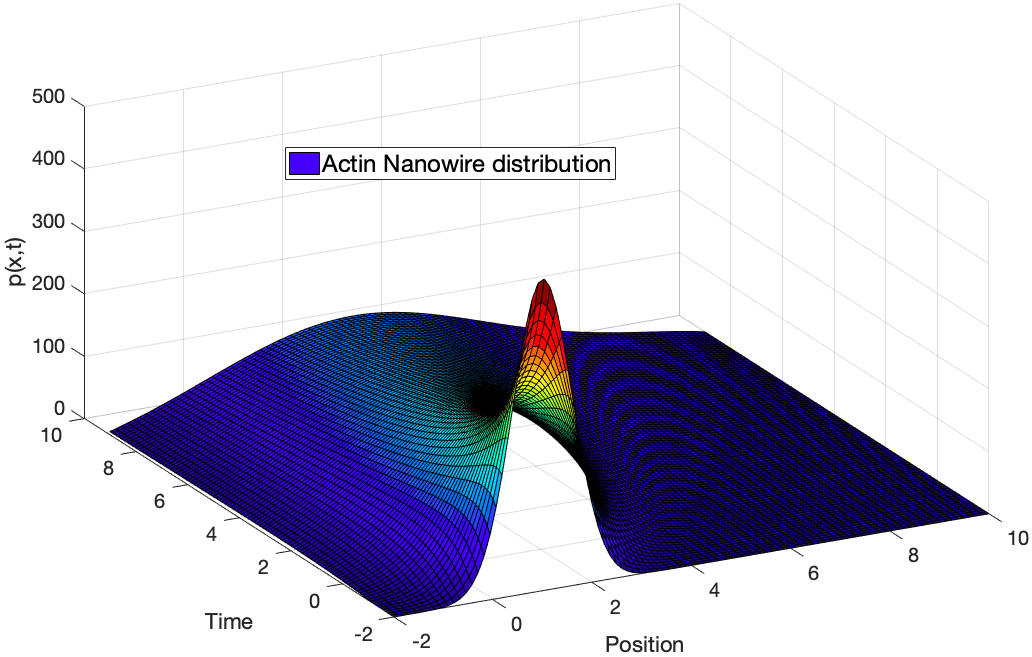}
		\caption{Probability density function of the actin nanowire distribution.}
		\label{fig:8}
	\end{figure}

	\begin{figure*}[!htb]
	\minipage{0.33\textwidth}
	\includegraphics[width=\columnwidth]{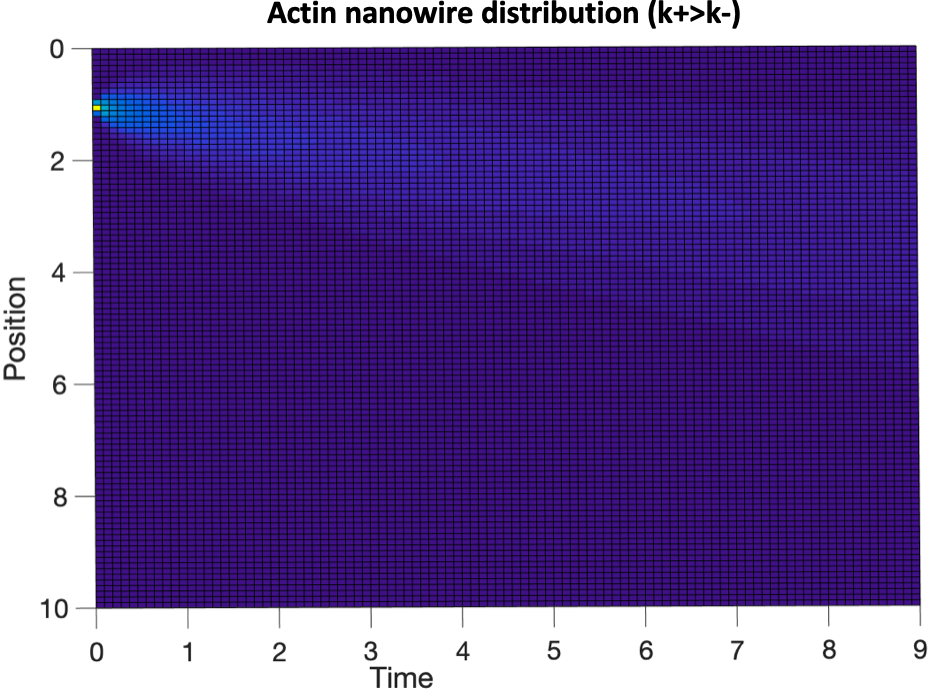}
	\endminipage\hfill
	\minipage{0.33\textwidth}
	\includegraphics[width=\linewidth]{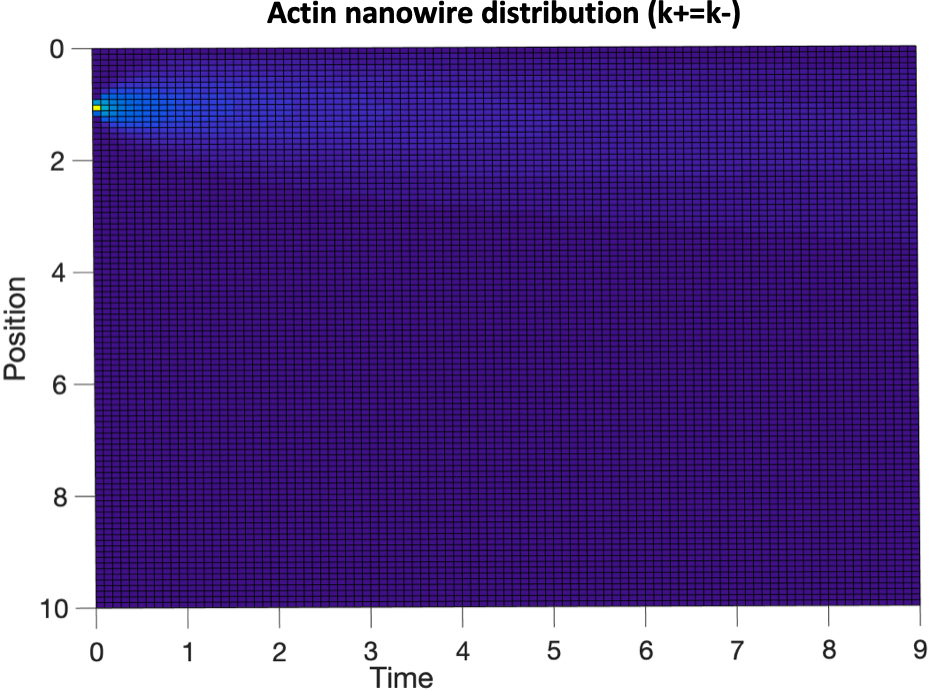}
	\endminipage\hfill
	\minipage{0.33\textwidth}
	\includegraphics[width=\linewidth]{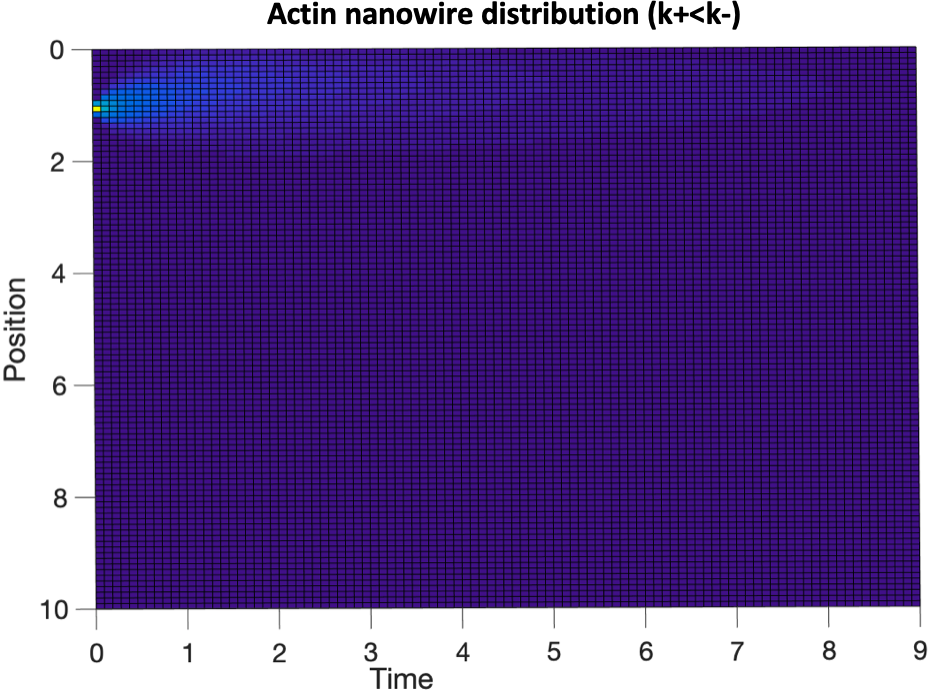}
	\endminipage
	\label{fig:9}
	\caption{The influence of the reaction rates on the actin nanowire distribution in three scenarios, $k_+ > k_-$, $k_+ = k_-$ and $k_+ < k_-$ respectively.}
\end{figure*} \begin{figure*}[!htb]
	\minipage{0.32\textwidth}
	\includegraphics[width=\columnwidth]{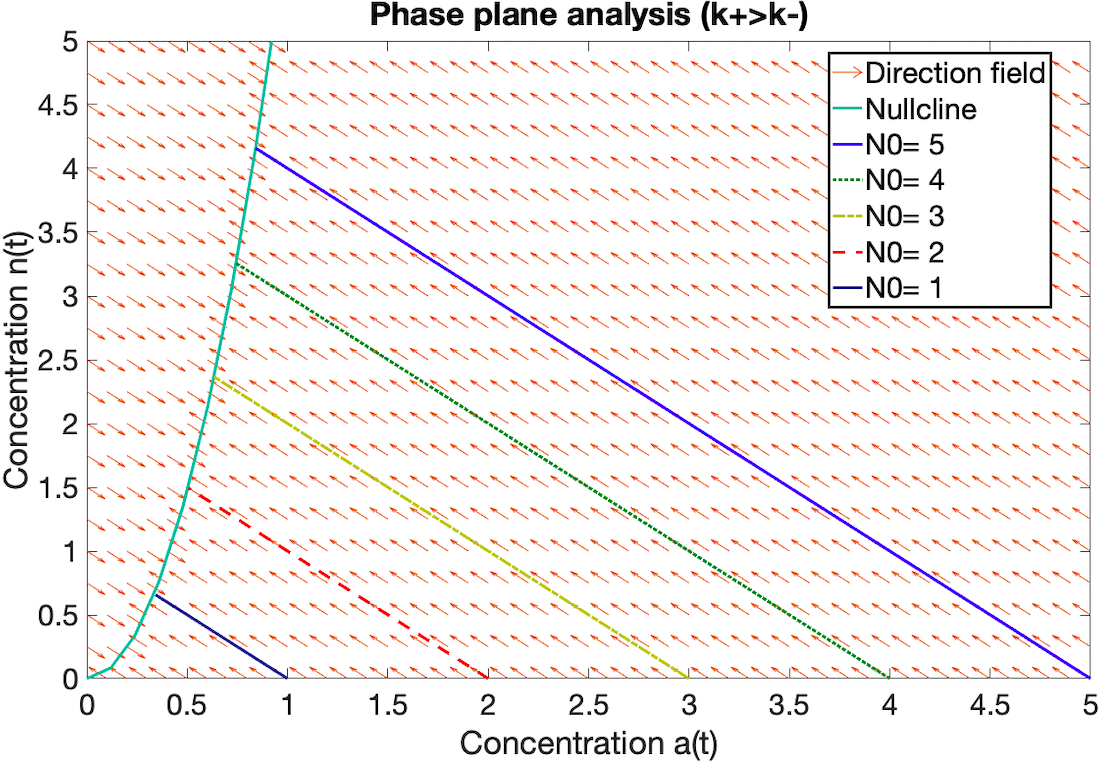}
	\endminipage\hfill
	\minipage{0.32\textwidth}
	\includegraphics[width=\linewidth]{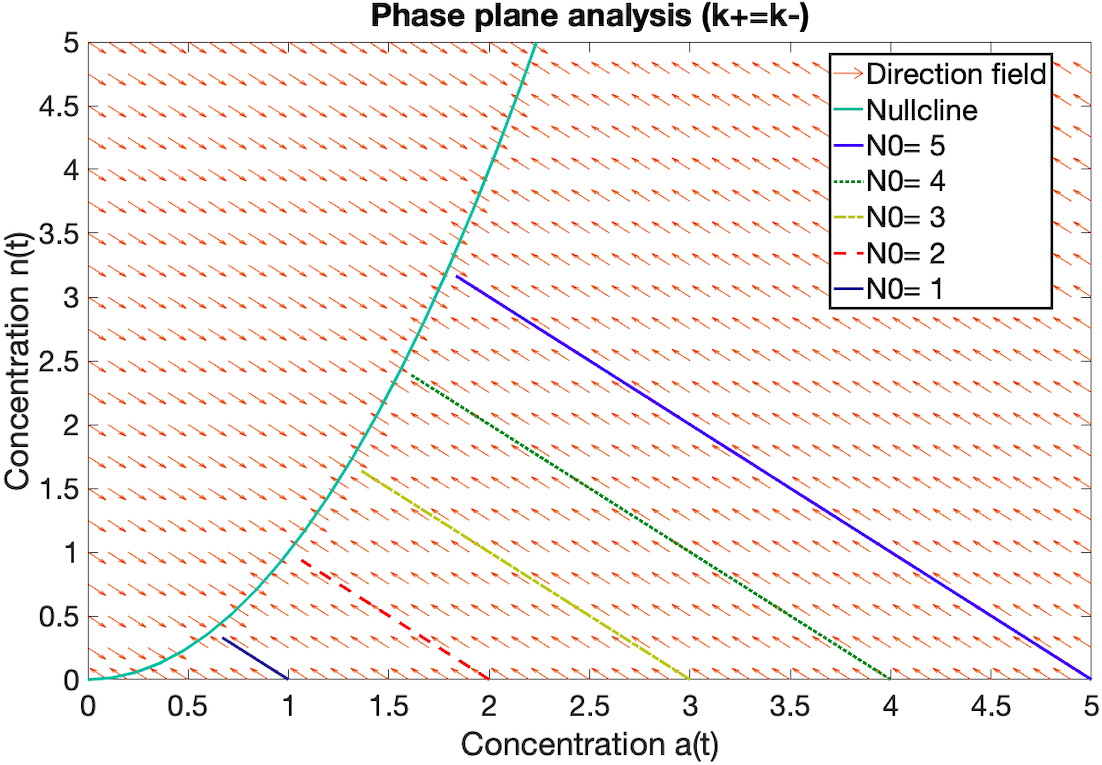}
	\endminipage\hfill
	\minipage{0.32\textwidth}
	\includegraphics[width=\linewidth]{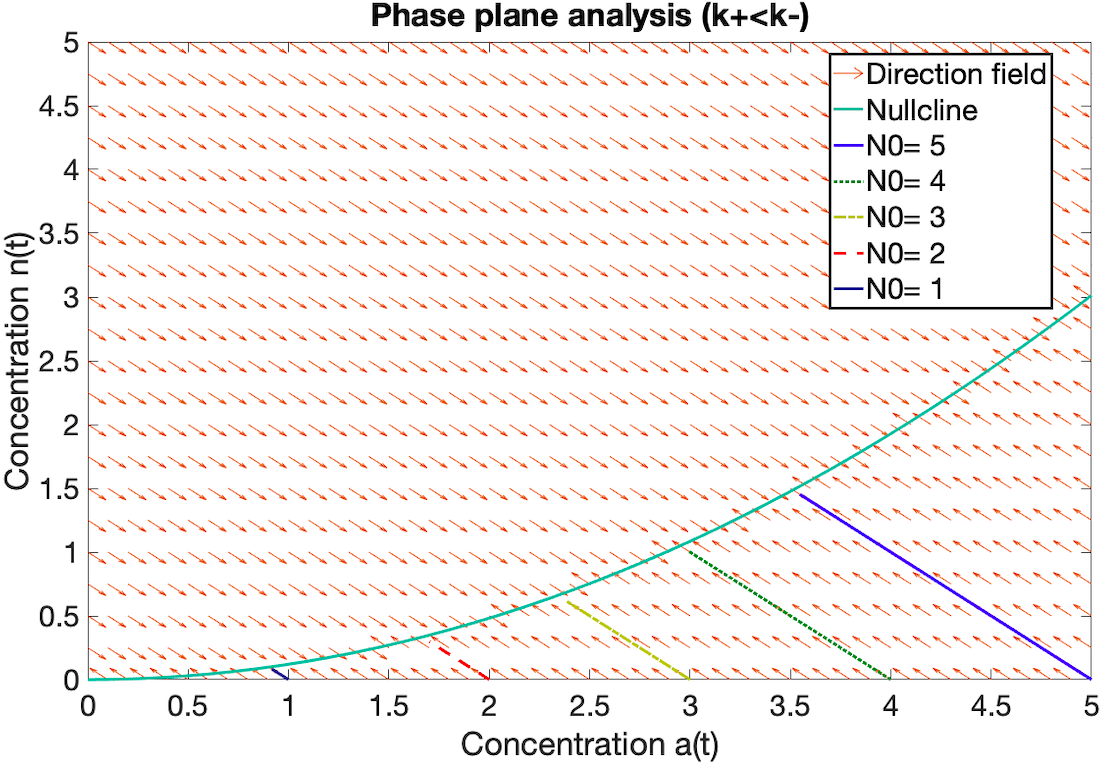}
	\endminipage
	\label{fig:10}
	\caption{Stability analysis of the actin nanowire formation under three scenarios,  $k_+ > k_-$, $k_+ = k_-$ and $k_+ < k_-$ respectively.}
\end{figure*}
	
	\subsection{Channel's Dynamics}
	
	The actin self-assembly chain reactions are very complex and to study their dynamics, we need to approximate the discrete changes in molecules' number into continues change in their concentration as shown in Fig. 7. The figure compares the analytical, numerical and Monte-Carlo simulation results of the system's differential equations described in (3) and (4), which represent the change in the actin monomer concentration while the actin nanowire elongates in the medium. We notice that the analytical solution calculated in (13) matches the numerical results and Monte-Carlo simulations. The steady states of the actin nanowire polymerization and actin monomer concentration are calculated from (12). Fig. 7 exhibits the dynamic behavior of the actin monomers in the medium and predicts the time needed for the nanowire to reach its steady state with the chosen initial conditions. Using the results of the derived model helps the user to determine the optimal distance channel for the nano-communication system.
	
	Fig. 8 shows the probability density function p(x,t) of the actin nanowire distribution. The numerical evaluation of the Fokker-Planck equation described in (22) and shown in Fig. 8 is simulated with MATLAB by using the derived coefficients of the diffusion and the elongation rate. We observe that the actin nanowire starts propagating from a position $x_0$ = 1  $\mu m$, where the first monomers are already anchored at the surface of the transmitter. Then, it elongates with a rate smaller than the diffusion coefficient, which explains the actin nanowire slow propagation. Most of the methods in the literature propose the use of the diffusion to propagate the information, whether by using the medium, bacteria or kinetic motors. In this study, we use the diffusion to elongate our proposed actin nanowire. But once it is attached to the receiver, the information is sent through it very rapidly by using electrons.

	Actin nanowire formation is greatly influenced by the reaction rates $k_+$ and $k_-$, which themselves are influenced by enzyme concentration, viscosity of the medium and magnetic field used to guide the direction of actin assembly. Fig. 9 shows the influence of the reaction rates on the actin nanowire distribution in three scenarios, $k_+ > k-$, $k_+= k_-$ and $k_+ < k_-$. We see that when $k_+ > k-$, the actin distribution is drifted from its initial position, which means that the actin nanowire starts assembling and its length doubled in 5 seconds. In the $k_+= k_-$ scenario, we notice that the position of the actin distribution does not change with time and stays at the initial position (1 $\mu m$). This is explained by the fact that the number of actin monomers added and dissociated from the nanowire are the same, thus, the nanowire does not elongate. The last scenario shows the collapse of the actin nanowire because the number of the dissociated monomers is bigger.

	The user of the actin-based nano-communication system can employ the reaction rates as a switch to enable or disable the nanowire formation, which gives more flexibility to the nanonetwork. Actin polymerization can also be inhibited by using cytochalasin enzymes \cite{fox_inhibition_1981}.

	\subsection{Stability Analysis of the Channel}
	
	Fig. 10 shows the influence of the reaction rates on the actin nanowire stability. To better analyze the stability of the nanowire, we plotted the concentrations against one another, instead of plotting them as functions of time. The phase plan graph in Fig. 10 plots the concentration n(t) of the actin monomers in the medium against the concentration a(t) of the actin monomers constructing the nanowire, in the three scenarios discussed above. To avoid that the phase plan becomes crowded with all possible trajectories, we plotted short arrows to indicate the motion direction, and we plotted five possible trajectories. The direction field is plotted to show the stability points of the proposed channel by using five initial concentrations of the actin monomers. We notice that whatever initial concentration we choose, the system follows the nullcline calculated in (15), which represents the equilibrium points of the system. As revealed in the derived nullcline expression, the stability of the channel depends on the reaction rates. The equilibrium points change in each scenario, but the system always follow these points and reaches its steady state.

	The stability of the channel depends on other parameters too, as explained in Fig. 6. The enzyme concentration, the length of the communication channel and the intensity of the magnetic field also affect the stability of the actin nanowire.

	\subsection{Channel Model Evaluation}
	
	The Probability density function of the actin nanowire distribution presented in Fig. 8 is dictated by a coin toss, either a molecule is added or dissociated from the nanowire. Depending on the random outcome of this probability, the actin nanowire elongates or collapses as explained by the dynamic behavior study of the proposed channel model. In order to evaluate the channel model, we compare the analytical solution of the Fokker-Planck equation calculated in (25) with the numerical simulations of the derived master equation in (10) by using Markov-Chain Monte-Carlo (MCMC) method. The comparison is shown in Fig. 11, where the nanowire starts distributing from its initial position ($x_0$ = 1 $\mu m$) and elongates towards the receiver position ($x_l$ = 10 $\mu m$) with time. We notice that the calculated analytical solution of the Fokker-Planck equation and the derived master equation simulations match favorably, which validates the proposed channel model. We observe that the nanowire reaches half the communication channel in 50 seconds (x= 5 $\mu m$) with the chosen parameters and initial conditions. The amplitude of p (x, t) that represents the actin monomers in the medium decreases with time because monomers are added to the elongated nanowire.
	
	\begin{figure}
		\centering
		\includegraphics[width=\linewidth]{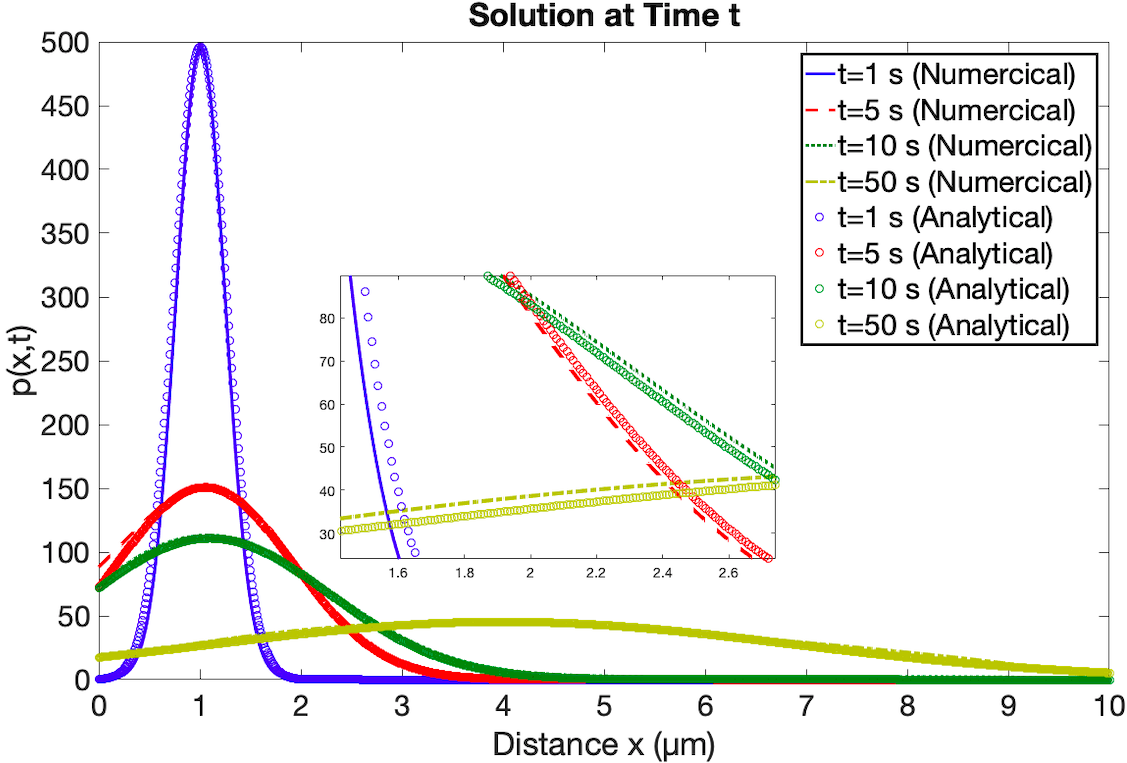}
		\caption{Comparison between the analytical solution of the Fokker-Planck equation and the numerical simulations of the channel's master equation.}
		\label{fig:11}
	\end{figure}
	
	\section{Conclusion}
	
	In this paper we modeled a nano-communication wired channel based on a polymer self-assembly. We studied the dynamic behavior of the channel by deriving the chemical master equation of the polymerization reaction. We formulated the proposed channel construction in two differential equations, we solved them analytically and we validated the solution with numerical and Monte-Carlo simulations. Then, we approximated the master equation by a one-dimensional Fokker-Planck equation and we solved it analytically and numerically. Moreover, we studied the nanowire stability and we derived the expressions of its diffusion coefficient and its elongation rate. 
	
	Using the numerical evaluation of the proposed model, we show that the reaction rates of the polymer assembly influence not only its distribution, but also its stability. The study also shows that the reaction rates can be used as a switch to enable or disable the nanowire formation by using enzymes, giving more flexibility to the polymer-based nanonetworks. In comparison with wireless molecular communication techniques proposed in the literature, the proposed wired polymer-based method promises stable and flexible nanonetworks with a much higher achievable throughput. Moreover, the proposed polymer-based nanonetworks are potentially biocompatible, which makes them a suitable candidate for designing bio-inspired nanonetworks for medical and pharmaceutical applications.
	
	\bibliographystyle{IEEEtran}
	\bibliography{ref}
	
\end{document}